\documentclass[12pt]{article}

\usepackage{amsmath,amsthm,a4wide}
\usepackage{epsfig}

\newcommand{\mycite}[1]{ \cite{#1}}

\newtheorem{lem}{Lemma}
\newtheorem{theo}{Theorem}
\numberwithin{equation}{section}

\title{Scaling behaviour of two-dimensional\\ polygon models}

\author{Christoph Richard\footnote{
Department of Mathematics and Statistics,
The University of Melbourne,
Parkville, Victoria 3010, Australia}
\footnote{
Institut f\"ur Mathematik und Informatik,
Universit\"at Greifswald,
Jahnstr.~15a, 17487 Greifswald, Germany (since January 2002)}}

\begin{document}

\maketitle

\begin{abstract}
Exactly solvable two-dimensional polygon models, counted by perimeter and area, are
described by $q$-algebraic functional equations.
We provide techniques to extract the scaling behaviour of these models up to arbitrary
order and apply them to some examples.
These are then used to analyze the unsolved model of self-avoiding polygons, 
where we numerically confirm predictions about its scaling function and its first 
two corrections to scaling.
\end{abstract}

$\mbox{ }$
\\

\section{Introduction}

Two-dimensional polygon models\mycite{V98, Ja00} such as self-avoiding 
polygons\mycite{LSF87,F89,FGW91}, counted by perimeter and area, 
are described in the grand-cononical ensemble where one
introduces a perimeter activity $x$ and an area activity $q$.
The perimeter and area generating function $G(x,q)$ is defined as
\begin{equation}
G(x,q) = \sum_{m,n} p_{m,n} x^m q^n = \sum_{m=0}^\infty p_m(q) x^m,
\end{equation}
where $p_{m,n}$ denotes the number of polygons on a given lattice 
with perimeter $m$ and area $n$.
The thermodynamical properties of polygons are described by the behaviour of the
perimeter and area generating function about its singular points.
The singular behaviour of the generating function translates to properties of the
coefficients $p_{m,n}$ in the limit of large perimeter or area. 
The locus of singularities (phase diagram) of a typical polygon model
is shown in Fig. 1.

\begin{figure}[htb]
  \begin{center}
    \leavevmode
    \epsfig{file=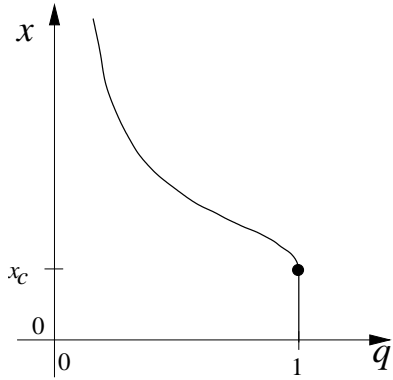,height=5cm}
    \caption{ Phase diagram of a typical two-dimensional polygon model}
    \label{fig:phase}
  \end{center}
\end{figure}

Of particular interest are points where lines of singularities meet,
which describe phase transitions.
For example, the model of self-avoiding polygons (SAPs) is believed to
exhibit tricritical behaviour, to be described below, about the point
where the perimeter generating function $G(x,1)$ is singular.
This point describes a crossover between extended and deflated 
self-avoiding polygons:
The mean area $\langle a \rangle_m$ of SAPs of perimeter $m$ grows 
asymptotically like\mycite{FGW91}
\begin{equation}
\langle a \rangle_m \sim \left\{ 
\begin{array}{ll} 
A(1) \, m^\frac{3}{2} & (q=1),\\ 
A(q) \, m & (0<q<1).
\end{array} \right.
\end{equation}
It can be shown\mycite{FGW91} that in the limit $q\to 0$ the generating
function is dominated by polygons of minimal area.
Since these polygons may be viewed as branched polymers\mycite{F89},
the phase $q<1$ is also referred to as the branched polymer phase.

Little exact information is known about SAPs\mycite{FGW91, CG93, C94,
Ja00}, almost all information being as a result of numerical 
studies\mycite{LSF87, FGW91, JG99, J00}.
Recently, a prediction of the functional form of the scaling function, 
which describes the tricritical behaviour, has been derived and confirmed 
numerically\mycite{RGJ01}.
This led, by re-interpretation using field-theoretic methods, to
the prediction of a number of other exact scaling functions\mycite{C01}.
In this paper, we explain our previous analysis\mycite{RGJ01} within a larger
mathematical framework and extend it to include analysis of corrections to scaling.

Due to the difficulty of the SAP problem, simpler polygon models, which
are subclasses of self-avoiding polygons, have been analyzed and
solved exactly\mycite{PO95b,Bou96,Ja00}.
Over the years, it has become clear that the underlying structure is described by 
$q$-algebraic functional equations\mycite{BG90b, L91, Bou92, BOP94, Bou96, Ja00}, 
which we will analyze in detail in this paper.
Assume that the perimeter and area generating function $G(x,q)$ of a polygon model
satisfies a $q$-algebraic functional equation\mycite{RGJ01}
\begin{equation} \label{form:func}
P( G(x,q), G(qx,q), \ldots, G(q^Nx,q),x,q) = 0,
\end{equation}
where $P(y_0,y_1,\ldots,y_N,x,q)$ is a polynomial in $y_0,y_1,\ldots,y_N,x$ and $q$.
The motivation for this type of equation arises from the requirement
that polygons of a given size can be built up recursively from polygons of
smaller size\mycite{BOP94, PB95, Bou96}.
The existence of such a recursion accounts for the solvability of the model.
We call polygon models of such a form $q$-algebraic.

The limit $q \to 1$ in (\ref{form:func}) leads to an {\it algebraic} differential equation
for the perimeter generating function $G(x,1)$ of order $N$, i.~e. a
vanishing polynomial in $x$, $G(x,1)$ and its derivatives.
From $q$-linear equations, the limit $q\to 1$ leads to a {\it linear} differential 
equation of order $N$.
This may be seen by first expressing $G(qx,q)$ in terms of the
$q$-derivative $(D_q G)(x) = (G(x,q)-G(qx,q))/(1-q)x$,
likewise $G(q^kx,q)$ in terms of higher $q$-derivatives, and then
expanding the functional equation about $q=1$.
The order $(1-q)^n$ in the expansion contains derivatives of degree not
exceeding $n$.
The defining equation for the perimeter generating function $G(x,1)$
is given by the non-trivial contribution of lowest order.
One simple case is given by $n=0$, where derivatives are absent.
In this case, the perimeter generating function satisfies the algebraic equation
\begin{equation} \label{form:algfunc}
P( G(x,1), G(x,1), \ldots, G(x,1),x,1) = 0.
\end{equation}
Generally, exactly solved polygon models have algebraic perimeter
generating functions\mycite{Bou96}.
However, it has been shown\mycite{R00} that the anisotropic perimeter generating 
function of the unsolved model of self-avoiding polygons is not algebraic 
(even not $D$-finite\mycite{S80}), and there are other simple, unsolved
polygon models such as three-choice polygons\mycite{CGD97}, which displays a
logarithmic singularity.
A natural question to ask is if these models are also described by $q$-algebraic 
functional equations.
As we showed\mycite{RGJ01}, for rooted self-avoiding polygons $G^{(r)}(x,q)=x\frac{d}{dx}G(x,q)$ 
this assumption appears to lead to the correct prediction for the singular part of the perimeter and
area generating function about the phase transition point, 
thereby answering a conjecture by Guttmann about the scaling function\mycite{PO95b}
in the affirmative.
Of course, this does not imply that rooted self-avoiding polygons are $q$-algebraic.

From physical grounds, we expect tricritical scaling behaviour about such points,
\begin{equation}\label{form:sclbeh}
G(x,q) \sim G^{(reg)}(x,q) + (1-q)^{\theta} F( (x_c-x) (1-q)^{-\phi}),
\qquad (x,q) \to (x_c^-, 1^-),
\end{equation}
where $F(s)$ is a scaling function, depending only on a combined
variable $s=(x_c-x)/(1-q)^{\phi}$, which is commonly assumed to be regular at the origin.
Its value at the origin determines the singular behaviour of the area
generating function $G^{(sing)}(x_c,q) \sim F(0)(1-q)^\theta$ about $q=1$.
The behaviour of the scaling function at infinity determines the
singular behaviour of the perimeter generating function $G(x,1)$ about
the critical point $x_c$.
If $F(s)\sim a s^{-\gamma}$ for $s \to \infty$, it follows that
$G(x,1)\sim a (x_c-x)^{-\gamma}$ for $x \to x_c$, if $\theta+\phi
\gamma=0$.
The exponent $\phi$ is called the crossover exponent.
The slope of the critical line $x_c(q)$ about $q=1$ is determined by
the first singularity $s_c$ of $F(s)$ on the negative real axis.
We have $x_c(q)\sim x_c-s_c(1-q)^\phi$ about $q=1$.
If $F(s)\sim a (s-s_c)^\alpha$ about $s=s_c$, the singular part of the
perimeter and area generating function behaves as $G^{(sing)}(x,q)\sim
a (1-q)^{-\alpha\phi} (x_c(q)-x)^\alpha$ for $x\to x_c(q)$.

From the scaling function of the grand-canonical ensemble, finite-size quantities may be
computed.
It can be shown\mycite{Ja00} that for large $m$ under mild assumptions
\begin{equation}
p_m(q) \sim m^{\theta-1} x_c^{-m-\theta} h \left(\frac{m}{x_c}
(1-q)^{-\phi} \right) \qquad (m \to \infty, q \to 1^-),
\end{equation} 
where the finite-size scaling function $h(x)$ is given by
\begin{equation}\label{form:fs}
h(x) = \sum_{k=0}^\infty \frac{f_k}{\Gamma (k\phi+\theta)} x^k,
\end{equation}
and the coefficients $f_k$ appear in the asymptotic expansion of the scaling function
\begin{equation}
F(s) = \sum_{k=0}^\infty \frac{f_k}{s^{(k-\theta)/\phi}}.
\end{equation}
Due to the appearance of the Gamma function in (\ref{form:fs}), the functional form of
the finite-size scaling function is generally more complicated than the 
grand-canonical scaling function. 
The determination of scaling functions using an exact solution is
known to be a difficult problem.
They have been extracted for semicontinuous versions of a number of
polygon models such as Ferrers diagrams and stacks\mycite{PO95a},
staircase polygons\mycite{PB95, P95} and column-convex polygons\mycite{BOP94}.
In some cases, explicit expressions for the finite-size scaling
function can be derived\mycite{OBE01}.

Recently it has been shown that the scaling function for simple
discrete models can be derived directly from the defining
$q$-functional equation\mycite{RG01,RGJ01}, and one is led to ask about 
the generality of the approach.
This is the aim of the present paper.
As a first step towards understanding the scaling behaviour of general
$q$-algebraic polygon models, we focus on the case where the perimeter
generating function is algebraic.
Remarkably, the generic form of the scaling behaviour of the perimeter
and area generating function is determined by the nature of the
algebraic singularity alone, 
irrelevant of the details of the functional equation.
This we illustrate in the simplest case of a square-root singularity and a simple
pole at $x=x_c$.
In these cases, we will perform a formal asymptotic expansion about
$x=x_c$ and $q=1$ which is uniform in $q$.
This asymptotic expansion does not only give the scaling behaviour but also gives
corrections to the scaling behaviour up to arbitrary order.
After presenting exactly solvable examples, we discuss the model of self-avoiding
polygons.
The perimeter generating function of {\it rooted} self-avoiding
polygons displays a square-root singularity, 
but is almost certainly not algebraic\mycite{R00}.
We will confirm numerically that the scaling function about the
singular point is the one predicted by the assumption of a 
$q$-algebraic functional equation for the perimeter and area
generating function.
We will also confirm predictions about the first two corrections to scaling.
The paper concludes with a discussion of open questions and possible future work. 

\section{Area moments of $q$-algebraic polygon models}

Let $G(x,q)$ be the perimeter and area generating function of a polygon model.
The area weighted moments $g_k(x)$ appear in the formal expansion about $q=1$
\begin{equation}
G(x,q) = \sum_{m,n} p_{m,n} x^m q^n = \sum_k g_k(x)(1-q)^k
\end{equation}
and are of the form
\begin{equation}
g_k(x) = \frac{(-1)^k}{k!} \sum_{m,n} n (n-1) \cdots (n-k+1) p_{m,n} x^m.
\end{equation}
Note that all area moments have the same radius of convergence as the perimeter 
generating function $g_0(x)$.
This can be seen\mycite{Ja00} by a ratio test, using the estimate 
\begin{equation}
n (n-1) \cdots (n-k+1) \le n^k \le m^{2k} \quad (n > n_0).
\end{equation}
The second inequality amounts to the fact that squared perimeter $m^2$ is larger than 
the area $n$ of a polygon. 

Let us assume that a polygon model satisfies the $q$-algebraic functional equation
\begin{equation}
P( y_0, y_1, \ldots, y_N,x,q) = 0, \qquad y_k = G(q^k x,q).
\end{equation}
We will show below that the functional equation determines the moments
$g_k(x)$ recursively.
In order to proceed, we have to specify the properties of the functional equation in the
limit $q \to 1$.
Therefore assume that the limit $q\to 1$ leads to the algebraic equation
\begin{equation}\label{form:algeq}
P( y_0, y_0, \ldots, y_0,x,1) = 0, \qquad y_0 = G(x,1).
\end{equation}
The singular behaviour of the perimeter generating function $G(x,1)$ is determined
by the algebraic equation (\ref{form:algeq})\mycite{H59}.
This in turn determines the singular behaviour of all other moments.
We will make this explicit for the two simplest forms of algebraic singularity: a
square-root singularity and a simple pole.
This generalizes previous investigations of the model of staircase
polygons, where singular exponents of the first two area moments have
been determined\mycite{PB95}. 

\subsection{Recursive determination of the area moments}

We show that the area weighted moments can be computed recursively from the functional
equation if the limit $q \to 1$ leads to a non-trivial algebraic equation for
the perimeter generating function $g_0(x)$.

It is instructive to first derive an expression for $g_1(x)$.
To this end, we expand the $q$-algebraic functional equation to first order in
$\epsilon=1-q$, making use of the expansion
\begin{equation}
G(q^kx,q) = g_0(x) + \epsilon (g_1(x) - k x g_0'(x)) + {\cal O}(\epsilon^2).
\end{equation}
The terms of order $\epsilon$ are given by
\begin{equation}\label{form:first}
g_1(x) \left( \sum_k \partial_k P \right) - x g_0'(x) \left( \sum_k k
\partial_k P \right) - \left( \partial_q P \right) =0.
\end{equation}
In the above equation, $P$ is evaluated at argument 
$(y_0,\ldots,y_N,x,q)=(y_0,\ldots,y_0,x,1)$, and we adopted the
notation $\partial_k=\partial_{y_k}$.
The first term is non-zero by the assumption of an algebraic equation for
$g_0(x)$ of the form (\ref{form:algeq}). 
This means that $g_1(x)$ is given in terms of powers of $g_0(x)$ and its derivative.
In the general situation the following statement holds:
\begin{lem}
{\rm The order $\epsilon^n$ in the expansion of the functional equation is linear in $g_n(x)$
and contains products of $g_k(x)$ and its derivatives with $0\le k<n$.}
\end{lem}
\begin{proof}
This is seen by an analysis of the terms appearing in the expansion of the functional
equation in powers of $\epsilon$.
We will do this in detail for subsequent purposes.
Note first that the expansion of a $q$-shifted function $f(q^kx)$ is given by
\begin{equation}\label{form:qshift}
f(q^k x) = 
\sum_{l=0}^\infty (-\epsilon)^l \binom{k x\partial_x}{l} f(x)=
\sum_{l=0}^\infty \epsilon^l \sum_{m=0}^l h(k,l,m) x^m f^{(m)}(x),
\end{equation}
where
\begin{equation}
h(k,l,m) = \frac{(-1)^l}{l!} \left( \sum_{r=m}^l S^{(1)}_{l,r} S^{(2)}_{r,m} k^r\right).
\end{equation}
Here, $S^{(1)}_{m,n}$ and $S^{(2)}_{m,n}$ denote the Stirling numbers of first and
second kind, respectively.
We have $h(k,0,0)=1$ and $h(k,l,m)=0$ for $m>l$.
The expansion of $G(q^kx,q)$ is then given by
\begin{equation}
G(q^kx,q) = g_0(x) + h_k(\epsilon,x) \qquad (k=0,\ldots,N).
\end{equation}
The function $h_k(\epsilon,x)$ has an expansion in powers of
$\epsilon$ of the form
\begin{equation}\label{form:shexp}
h_k(\epsilon,x) = \sum_{n=1}^\infty \epsilon^n h_{k,n}(x), \qquad 
h_{k,n}(x) = \sum_{l=0}^n \sum_{r=0}^l h(k,l,r) x^r g_{n-l}^{(r)}(x).
\end{equation}
With the introduction of
\begin{equation}
H = (h_0(\epsilon,x), \ldots, h_N(\epsilon,x), -\epsilon), \qquad
\nabla = (\partial_0, \ldots, \partial_N, \partial_q),
\end{equation}
the expansion of the $q$-functional equation can be written as
\begin{equation}
\begin{split}
P(y_0, & \ldots, y_N,x,q) = \sum_{k=0}^\infty \frac{(H \nabla)^k}{k!}
P(y_0,\ldots,y_0,x,1)\\
&=\sum_{r,m=0}^\infty \sum_{{\bf k}_m}
\frac{(-\epsilon)^r}{m! \, r!}h_{k_1}(\epsilon,x) \cdots h_{k_m}(\epsilon,x) 
\partial_q^r \partial_{{\bf k}_m} P(y_0,\ldots,y_0,x,1).
\end{split}
\end{equation}
Note that the first sum is finite since $P(y_0,\ldots,y_N,x,q)$ is
assumed to be a polynomial in its variables.
In the second equation we introduced the abbreviation ${\bf
x}_n=x_1,\ldots,x_n$.
The range of summation is $0\le k_i\le N$.
We now insert (\ref{form:shexp}) into the above equation and extract
the term of order $\epsilon^n$.
It is given by
\begin{equation}\label{form:epscont}
\begin{split}
\sum_{r,m=0}^\infty \frac{(-1)^r}{m! \, r!} 
\sum_{{\bf k}_m} 
\sum_{
\genfrac{}{}{0pt}{1}{|{\bf r}_m|}{=n-r}
}
\sum_{
\genfrac{}{}{0pt}{1}{{\bf s}_m}{{\bf t}_m}
}
x^{|{\bf t}_m|} 
& \left( \prod_{i=1}^m h(k_i,s_i,t_i) g_{r_i-s_i}^{(t_i)}(x) \right) \cdot\\
& \cdot \partial_q^r \partial_{{\bf k}_m} P(y_0,\ldots,y_0,x,1),
\end{split}
\end{equation}
where we introduced the abbreviation $|{\bf x}_n|=x_1+\cdots+x_n$.
We have $1\le r_i$, $0 \le s_i \le r_i$ and $0 \le t_i \le s_i$.
The summations over ${\bf k}_m$ and ${\bf r}_m$ are independent.
It follows from (\ref{form:epscont}) that the order $\epsilon^n$ contains
$g_n(x)$ only linearly:
The constraint $|{\bf r}_m|=n-r$ implies $r=0$ and $m=1$.
This in turn implies $s_1=t_1=0$ and leads to
$g_n(x) \sum_{k} \partial_{k} P(y_0,\ldots,y_0,x,1)$.
As we argued above, this term is non-zero.
The other terms are given by products of derivatives of $g_k(x)$ with $k<n$.
It is therefore possible to solve for the area moments recursively.
\end{proof}

\subsection{Behaviour about a square-root singularity}

Assume that the algebraic perimeter generating function 
$G(x,1)$ has a square-root singularity at $x=x_c$.
It follows from the theory of algebraic functions\mycite{H59} that this
is equivalent to the conditions
\begin{equation}\label{form:sqcon1}
\left( \sum_k \partial_k P \right) = 0, \qquad 
\left( \sum_{k,l} \partial_{k,l} P \right) \neq 0, \qquad 
\left( \partial_x P \right) \neq 0,
\end{equation}
on the functional equation, evaluated at $x=x_c$.
We have introduced the abbreviations $\partial_k=\partial_{y_k}$ and 
$\partial_{k,l}=\partial_{y_k}\partial_{y_l}$ for $0\le k,l \le N$.
The first condition implies that $x=x_c$ is a singular point, the
second and third condition indicate a square-root singularity.
The expansion is of the form
\begin{equation}\label{form:sqroot}
g_0(x) = g_0(x_c) + \sum_{l=0}^\infty f_{0,l}(x_c-x)^{\frac{l+1}{2}}.
\end{equation}
The coefficients $f_{0,l}$ can be determined recursively from the algebraic equation
(\ref{form:algfunc}), using the constraints (\ref{form:sqcon1}).
We have in particular
\begin{equation}\label{form:f00}
f_{0,0}^2 = \frac{\left(\partial_x P\right)}{\frac{1}{2} \left(\sum_{k,l}
\partial_{k,l} P\right)}.
\end{equation}
The sign of $f_{0,0}$ determines the branch of the solution.
Since we are dealing with generating functions having non-negative Taylor coefficients,
the negative sign has to be chosen.
With the expressions given above it is easy to analyze the singular behaviour of
$g_1(x)$.
To this end, we differentiate the algebraic equation (\ref{form:algfunc}) in order to 
replace $\left( \sum_k \partial_k P \right)$ in (\ref{form:first}).
We get
\begin{equation}\label{form:firstsq}
g_1(x) = - \frac{\left( \sum_k k \partial_k P \right)}{\left( \partial_x P \right)} x 
\left( g_0'(x) \right)^2 - \frac{\left( \partial_q P \right)}{\left(\partial_x P\right)} 
g_0'(x).
\end{equation}
We now make the assumption that
\begin{equation}\label{form:sqcon2}
\left( \sum_k k \partial_k P \right) \neq 0.
\end{equation}
This generic case is satisfied in the models discussed below.
It then follows that $g_1(x)$ admits a series expansion of the form
\begin{equation}
g_1(x) =  \sum_{k=0}^\infty f_{1,k}(x_c-x)^{\frac{k}{2}-1},
\end{equation}
where the coefficients $f_{1,k}$ can be computed from (\ref{form:firstsq}).
The leading coefficient $f_{1,0}$ is explicitly given by
\begin{equation}\label{form:f10}
-4 f_{1,0} = x_c 
\frac{\left( \sum_k k \partial_k P \right)}{\frac{1}{2}\left(
\sum_{k,l} \partial_{k,l} P \right)}.
\end{equation}
For moments of higher order, we have the following result:
\begin{theo}{\rm
\item Under the assumptions (\ref{form:sqcon1}) and (\ref{form:sqcon2}), the area 
moment $g_n(x)$ admits an expansion about its singular point $x_c$ of the form
\begin{equation}
g_n(x) = \sum_{k=0}^\infty f_{n,k} (x_c-x)^{\frac{k}{2}-\frac{3}{2}n + \frac{1}{2}}.
\end{equation}
It can be computed recursively from the $q$-algebraic functional equation at order
$\epsilon^n$.
The coefficients $f_{n,0}$ are explicitly given by
\begin{equation}\label{form:sqas1}
f_{n,0} = c_n f_{1,0}^n f_{0,0}^{1-n},
\end{equation}
where the coefficients $f_{1,0}$ and $f_{0,0}$ are given by (\ref{form:f10}) and
(\ref{form:f00}), and $c_n$ satisfies the quadratic recursion relation
\begin{equation}\label{form:sqas}
c_n + (3n-4) c_{n-1} + \frac{1}{2}\sum_{r=1}^{n-1} c_{n-r}c_r, \qquad c_0=1.
\end{equation}}
\end{theo}
$\mbox{ }$\\
{{\bf Remarks:}\\
(1) The behaviour of the leading coefficients is independent of the degree $M \ge 2$ of 
the $q$-algebraic functional equation.\\
(2) The first few values of the coefficients are
$c_1=1$, $c_2=-5/2$,
$c_3= 15$, $c_4=-1105/8$, $c_5=1695$, $c_6=-414125/16$, $c_7=472200$, 
$c_8=-1282031525/128$, $c_9=242183775$, $c_{10}=-1683480621875/256$. \\ 
}
\begin{proof}
We first prove the stated form of the expansion of the functions $g_n(x)$ about the
singular point.
This we do by induction.
Since the case $n=1$ has been treated above, assume now that the statement holds for 
$g_k(x)$ where $k<n$.
We analyze the $q$-algebraic functional equation at order $\epsilon^n$.
By Lemma 1, it contains $g_n(x)$ linearly, and the other terms contain
products of derivatives of $g_k(x)$ with $k<n$.
From these facts it follows that $x=x_c$ is a singular point of $g_n(x)$ and that there
exists an expansion of $g_n(x)$ about $x=x_c$ in half-integer exponents.
To get the leading singular exponent of $g_n(x)$, we analyze the singular 
behaviour in a typical term of the form (\ref{form:epscont}), 
which is determined by the product appearing there.
Terms not containing $g_n(x)$ have by induction assumption a negative exponent of modulus
\begin{equation}
\frac{3}{2}\left( n-r \right) - \frac{m}{2} + 
\sum_{i=1}^m \left( t_i - \frac{3}{2} s_i \right).
\end{equation}
The leading singular behaviour is obtained by its maximum value
subject to the constraints $0\le t_i\le s_i$.
The maximum value appearing is $3n/2-1$.
It is realized by $g_{n-1}'(x)$ with values $(r,m,t_1,s_1)=(0,1,1,1)$
or by terms quadratic in $g_k(x)$ with values
$(r,m,t_i,s_i)=(0,2,0,0)$.
This leads to
\begin{equation}
\begin{split}
-g_n(x) & \frac{\left( \partial_x P \right)}{g_0'(x)} 
- \left( \sum_k k \partial_k P \right) x g_{n-1}'(x)
+ \frac{1}{2} \left( \sum_{k,l} \partial_{k,l} P \right) \sum_{r=1}^{n-1} g_{n-r}(x) g_r(x) \\
& = {\cal O}( (x_c-x)^{\frac{3}{2}-\frac{3}{2}n}).
\end{split}
\end{equation}
We can read off the leading singular behaviour of $g_n(x)$ from this equation.
Since the prefactors appearing are all non-vanishing, we get an exponent of $(1-3n)/2$.
The above equation also determines the values of the leading coefficients
(\ref{form:sqas}), as is seen by a straightforward computation.
\end{proof}

\subsection{Behaviour about a simple pole}

In this case it is advantageous to rewrite the $q$-algebraic functional equation in the form
\begin{equation}
P(y_0,\ldots,y_N,x,q) = \sum_{m=1}^M \sum_{{\bf k}_m} 
a_{{\bf k}_m}(x,q) y_{k_1} \cdots y_{k_m} +  a_0(x,q).
\end{equation}
We introduce the abbreviation $a_m(x,q) = \sum_{{\bf k}_m} a_{{\bf
k}_m}(x,q)$ and $a_m(x)=a_m(x,1)$.
In the limit $q\to 1$, the perimeter generating function $g_0(x)$ satisfies the algebraic
equation
\begin{equation}
a_M(x) g_0(x)^M + a_{M-1}(x) g_0(x)^{M-1} + \cdots + a_1(x) g_0(x) + a_0(x) = 0.
\end{equation}
We assume that at the critical point $x=x_c$ the algebraic equation satisfies
\begin{equation}\label{form:p1}
a_M(x_c) = 0, \qquad a_{M}'(x_c) \neq 0, \qquad a_{M-1}(x_c) \neq 0. 
\end{equation}
This is equivalent to the assumption of a simple pole.
From the theory of algebraic functions\mycite{H59} it follows that
$g_0(x)$ has an expansion of the form
\begin{equation}
g_0(x) = \sum_{k=0}^\infty f_{0,k} (x_c-x)^{k-1},
\end{equation}
where the coefficients $f_{0,k}$ can be calculated from the algebraic equation.
In particular, the first coefficient is given by
\begin{equation}\label{form:pf0}
f_{0,0} = \frac{a_{M-1}}{a_M'}(x_c).
\end{equation}
An expansion of the $q$-algebraic functional equation to first order in $\epsilon$ gives
\begin{equation}\label{form:fsimp}
\begin{split}
g_1(x) & \left( \sum_{m=1}^M m a_m(x) g_0(x)^{m-1} \right)
- \left( \sum_{m=0}^M \left(\partial_q a_{m}\right) (x,1) g_0(x)^m \right)\\
& - g_0'(x) \left( x \sum_{m=1}^M \left( \sum_{{\bf k}_m} |{\bf k}_m| a_{{\bf
k}_m}(x,1)\right) g_0(x)^{m-1}\right) =0.
\end{split}
\end{equation}
This is exactly (\ref{form:first}), but the present notation is better
suited to analyze the singular behaviour of $g_1(x)$, since the
divergent behaviour of the perimeter generating function $g_0(x)$ has
to be taken into account.
The above equation shows that the first area moment has an expansion about 
$x=x_c$ in integer powers.
We assume that
\begin{equation}\label{form:p2}
 \left( \sum_{{\bf k}_M}|{\bf k}_M| a_{{\bf k}_M}\right) (x_c,1) \neq 0.
\end{equation}
This generic behaviour is satisfied for the model discussed below.
The leading singular term in the equation is then of order $(x_c-x)^{-M-1}$,
and we conclude that $g_1(x)$ has an expansion of the form
\begin{equation}
g_1(x) = \sum_{k=0}^\infty f_{1,k} (x_c-x)^{k-3}.
\end{equation}
The expansion coefficients can be computed from the above equation.
The first coefficient is given explicitly by
\begin{equation}\label{form:pf1}
- f_{1,0} = x_c 
\frac{\left(\sum_{{\bf k}_M}|{\bf k}_M| a_{{\bf k}_M}\right)}{a_M'} 
\frac{a_{M-1}}{a_M'}.
\end{equation}
All expressions are evaluated at argument $(x,q)=(x_c,1)$.
In the general case, the following statement holds:
\begin{theo}{\rm
Under the assumptions (\ref{form:p1}) and (\ref{form:p2}), the area moment 
$g_n(x)$ admits an expansion about its singular point $x_c$ of the form
\begin{equation}
g_n(x) = \sum_{k=0}^\infty f_{n,k} (x_c-x)^{k-2n-1}.
\end{equation}
It can be computed recursively from the $q$-algebraic functional equation at order
$\epsilon^n$.
The coefficients $f_{n,0}$ are given by
\begin{equation}
f_{n,0} = c_n f_{1,0}^n f_{0,0}^{1-n},
\end{equation}
where $f_{0,0}$ and $f_{1,0}$ are given by (\ref{form:pf0}) and (\ref{form:pf1}), 
respectively.
The numbers $c_n$ satisfy the linear recursion relation
\begin{equation}\label{form:polrec}
c_n + (2n-1) c_{n-1} = 0, \qquad c_0=1.
\end{equation}}
\end{theo}
$\mbox{ }$\\
{{\bf Remarks:}\\
(1) The result is independent of the degree $M>0$ of the functional equation.\\
(2) The coefficients are given explicitly by $c_n=(-1)^n \frac{(2n-1)!}{2^{n-1}(n-1)!}$.
The first few values are $c_1=-1$, $c_2=3$, $c_3=-15$, $c_4=105$, $c_5=-945$, 
$c_6=10395$, $c_7=-135135$, $c_8=2027025$.\\ 
}
\begin{proof}
This statement is proved similarly to the square-root case treated above.
Assume that the statement holds for $g_k(x)$ where $k<n$.
We analyze the singular behaviour of the contribution from order
$\epsilon^n$.
As follows from Lemma 1, the point $x=x_c$ is a singular point of 
$g_n(x)$ and allows for an expansion of $g_n(x)$ in integer exponents.
Now consider terms containing $g_{n-1}'(x)$.
The prefactor appearing is exactly the second bracket in
(\ref{form:fsimp}).
Using the induction assumption, this results in a divergence with
exponent $2n+M-1$.
This is also the leading divergence:
Consider the singular behaviour in a general term in the sum
(\ref{form:epscont}).
Using the induction assumption again, we see that the product appearing
there diverges with an exponent of absolute value
\begin{equation}\label{form:polsingex}
2 \left( n-r \right) +m + \sum_{i=1}^m \left( t_i - 2 s_i \right).
\end{equation}
The derivatives of $P$ in (\ref{form:epscont}) contain diverging terms as well.
It can be computed that the most singular contribution is $g_0(x)^{M-m}$.
If $s_i=t_i=0$, this term is accompanied by $a_M$, 
adding an exponent $M-m-1$ to (\ref{form:polsingex}), due to the pole condition.
Otherwise, the divergence adds an exponent $M-m$.
This leads, by power counting, to two different choices of exponents:
$(r,m,t_i,s_i)=(0,m,0,0)$ or $(0,m,1,1)$ for exactly one $i$.
An explicit calculation of the prefactors appearing yields the
functional equation to leading order
\begin{equation}
\begin{split}
& \sum_{m=1}^M  
\left( \sum_{|{\bf r}_m|=n} g_{r_1}(x)\cdots g_{r_m}(x)\right)\cdot\\
\cdot & \left( -\binom{M}{m} g_0(x)^{M-m} a_M' +  
\binom{M-1}{m} g_0(x)^{M-1-m} a_{M-1}\right)\\
& - x \sum_{m=1}^M 
\left( \sum_{|{\bf r}_m|=n} g_{r_1-1}'(x)g_{r_2}(x)\cdots g_{r_m}(x)
\right)
\cdot \\
\cdot & 
\left( \sum_{{\bf k}_M}|{\bf k}_M| a_{{\bf k}_M} \right)
\binom{M-1}{m-1} 
g_0(x)^{M-m} = {\cal O} \left( (x_c-x)^{2-2n-M} \right).
\end{split}
\end{equation}
An inspection of the $m=1$ term reveals a divergence with exponent 
$2n+1$ for $g_n(x)$, as claimed above.
From the above equation a recursion relation for the values of the leading coefficients 
can be obtained.
It is
\begin{equation}
\begin{split}
 \sum_{m=1}^M \binom{M-1}{m-1} &
\left[ \left( \sum_{|{\bf r}_m|=n} c_{r_1}\cdots c_{r_m} \right) + \right.\\
&+\left. \left( \sum_{|{\bf r}_m|=n} (2 r_1-1) c_{r_1-1}c_{r_2}\cdots c_{r_m} \right)
\right] =0.
\end{split}
\end{equation}
This equation determines the values $c_n$ recursively, which may be seen by 
induction on $n$, using $r_i>0$.
We get the {\it linear} recursion (\ref{form:polrec}), being independent on $M$.
\end{proof}

\subsection{Area moments and scaling behaviour}

We now explain how the singular behaviour of the area moments is
related to the scaling function and the corrections to scaling.
Assume that the expansion of the area moments $g_k(x)$ about the singular point 
$x=x_c$ is given by
\begin{equation} \label{form:momex}
g_k^{(sing)}(x) = \sum_{l=0}^\infty \frac{f_{k,l}}{(x_c-x)^{\gamma_{k,l}}},
\end{equation}
where $\gamma_{k,l+1}<\gamma_{k,l}$, and $f_{k,l}$ are the amplitudes.
We next assume that the exponents $\gamma_{k,l}$ are of the special form
\begin{equation}
\gamma_{k,l} = \frac{k}{\phi} -\frac{\theta_l}{\phi}.
\end{equation}
This implies that $\theta_{l+1}>\theta_l$ and means that we can order the singular 
terms columnwise, where in each column the exponent increases by a constant 
indexed by $k$.
In this special situation, which is satisfied in our examples, the
singular part of the perimeter and area generating function has a formal expansion of the form
\begin{equation}
G^{(sing)}(x,q) = \sum_{l=0}^\infty \epsilon^{\theta_l} 
F_l\left( (x_c-x)\epsilon^{-\phi}\right),
\end{equation}
where $F_0(s)$ is the scaling function (\ref{form:sclbeh}), 
and $F_l(s)$ for $l>1$ are the correction-to-scaling functions.
We note scaling exponents $\theta_0$ and $\phi$ may be obtained 
from the analysis of the leading singular behaviour of the first two
area moments.
This technique was used intensively before\mycite{BOP94, PB95}.
The amplitudes $f_{k,l}$ appear in the asymptotic expansion of the
functions $F_l(s)$ via
\begin{equation}
F_l(s) = \sum_{k=0}^\infty \frac{f_{k,l}}{s^{\gamma_{k,l}}}.
\end{equation}
In the next section we will derive differential equations for the
scaling and correction-to-scaling functions $F_l(s)$ directly from the functional equation.

\section{Scaling of $q$-algebraic functional equations}

Assume that the perimeter generating function has an algebraic singularity at $x=x_c$. 
Assume further that the singular part of the perimeter and area generating function 
displays scaling behaviour about $x=x_c$ and $q=1$ of the form
\begin{equation}\label{form:scalbeh}
G(x,q) \sim G^{(reg)}(x,q) + (1-q)^{\theta} F_0\left( (x_c-x)
(1-q)^{-\phi}\right) \quad (x,q) \to (x_c^-,1^-).
\end{equation}
We expand the functional equation in terms of the
scaling function $F_0(s)$ at argument $s=(x_c-x) (1-q)^{-\phi}$.
In the limit $q \to 1$, this will lead to a differential equation for the scaling function,
as has been demonstrated for a number of simple examples previously\mycite{RG01, RGJ01}.
The technique is inspired by the method of dominant balance, which has been used
to derive scaling functions for semicontinuous versions of polygon
models\mycite{OPB93,BOP94,PB95}.
Introducing $\epsilon=1-q$, we first give the expansion of a
$q$-shifted function in analogy to (\ref{form:qshift})
\begin{equation}\label{form:qshift2}
\begin{split}
F & \left((x_c-q^k x) \epsilon^{-\phi}\right) = 
\sum_{l=0}^\infty (-\epsilon)^l 
\binom{k (s-x_c\epsilon^{-\phi})\partial_s}{l} F(s) \\
& = \sum_{m=0}^\infty F^{(m)}(s) \sum_{n=0}^m\ (-x_c)^n \binom{m}{n}
s^{m-n} \sum_{l=m}^\infty \epsilon^{l-n\phi} h(k,l,m).
\end{split}
\end{equation}
This implies that the $m$-th derivative scales with an exponent
$l-n\phi$ in $\epsilon$, where $0\le n \le m \le l$.
The smallest exponent occurs for $l=n=m$.
In the expansion of the scaling function of $q$-shifted argument this
leads to $m$-th derivatives $F_0^{(m)}(s)$ with scaling exponent
$\theta+m(1-\phi)$.
If $\phi>1$, this restricts the scaling function to be a polynomial in $s$:
In the expansion of the functional equation, derivatives of the scaling 
function will appear with arbitrarily small exponents in $\epsilon$.
This means that the limit $\epsilon \to 0$ can only exist if almost all 
derivatives of the scaling function vanish, i.e. if the scaling function is a polynomial.
We assume in the following that $\phi<1$.
This assumption is satisfied for the cases treated below and will result in a 
differential equation of first order for the scaling function.
If $\phi=1$, higher derivatives cannot be ignored, leading to a difference equation for
the scaling function\mycite{RG01}.
The expansion of the generating function with $q$-shifted argument is given to lowest order by
\begin{equation}\label{form:scalex}
G(q^k x,q) = G^{(reg)}(x,q) + \epsilon^{\theta} F_0(s) 
+ k x_c \epsilon^{\theta+ (1-\phi)} F_0'( s) + 
{\cal O}( \epsilon^{\theta+ 2(1-\phi)}).
\end{equation}
We now analyze the case of a square-root singularity at $x=x_c$ in detail.

\subsection{Scaling function about a square-root singularity}

We first derive a differential equation for the scaling function about a square-root
singularity and solve it.
In order to keep notation simple, we omit arguments, 
which are understood to be always at $(y_0,\ldots,y_N,x,q)=(y_0,\ldots,y_0,x_c,1)$.
Using (\ref{form:scalex}), we expand the functional equation
(\ref{form:func}) in powers of $\epsilon$.
We assume a square-root singularity, which implies $\phi=2\theta$ and
\begin{equation}
P = 0, \qquad 
\left( \sum_k \partial_k P \right) = 0, \qquad 
\left( \sum_{k,l}\partial_{k,l} P \right) \neq 0, \qquad 
\left( \partial_x P \right) \neq 0.
\end{equation}
The lowest orders in $\epsilon$ are given by
\begin{equation}\label{form:scaldglsqr}
\begin{split}
0 &= \epsilon^{2\theta} \frac{1}{2}  \left( \sum_{k,l} \partial_{k,l} P \right)  F_0^2(s)
-  \epsilon^{\phi} s \left( \left( \partial_x P \right) +
\left( \sum_k \partial_k P \right) \left(\partial_x G^{(reg)}\right) \right) \\
& + \epsilon^{\theta+(1-\phi)}  x_c \left(\sum_k k P_k \right) F_0'(s) +
\cdots
\end{split}
\end{equation}
The assumption that the two exponents do not coincide leads to a trivial scaling
function.
In order to have a nontrivial scaling function, we therefore demand 
$2\theta=\phi=\theta+(1-\phi)$ 
and get exponents
\begin{equation}
\theta = \frac{1}{3}, \qquad \phi=\frac{2}{3}.
\end{equation}
This coincides with the derivation using the area moments.
We obtain a differential equation for the scaling function of the form
\begin{equation}\label{form:diffeq}
F_0(s)^2 - 4 f_{1,0} F_0'(s) - f_{0,0}^2 s = 0,
\end{equation}
where the coefficients $f_{1,0}$ and $f_{0,0}$ are given by
\begin{equation}
f_{0,0}^2 = \frac{\left( \partial_x P \right)}{\left( \sum_{k,l}
\partial_{k,l} P \right)},  \qquad
f_{1,0} = - \frac{x_c}{2} \frac{\left(\sum_k k \partial_k P \right)}
{ \left(\sum_{k,l} \partial_{k,l} P\right)}.  
\end{equation}
The solution is uniquely determined by the prescribed asymptotic
behaviour at infinity and is given by
\begin{equation}\label{form:logairy}
F_0(s) = -4 f_{1,0} \frac{d}{ds} \ln \mbox{Ai} \left( \left(\frac{f_{0,0}}{4
f_{1,0}}\right)^{2/3}s\right),
\end{equation}
where $\mbox{Ai}(x)=\frac{1}{\pi}\int_0^\infty\cos(t^3/3+tx) \, dt$ is the Airy function.
The coefficients $f_{k,0}$ in the asymptotic expansion of the scaling function
are exactly those derived in the previous section.
The asymptotic behaviour is given by
\begin{eqnarray}
F_0(s) &\sim& f_{0,0} \, s^{1/2} \qquad (s\to \infty)\\
F_0(0) &=& \frac{1}{\pi}f_{1,0} \left( \frac{f_{0,0}}{4f_{1,0}} \right)^{2/3}
\Gamma\left(\frac{2}{3}\right)^2 3^{5/6}\nonumber
\end{eqnarray}
The first singularity on the negative real axis is a simple pole.
It is located  at 
$\left( \frac{f_{0,0}}{4f_{1,0}} \right)^{2/3}s_c=-2.338107\ldots$.

\subsection{Corrections to scaling about a square-root singularity}

The analysis of the singular behaviour of the area weighted moments
justifies a formal scaling Ansatz
\begin{equation}
G^{(sing)}(x,q) = \sum_{l=0}^\infty \epsilon^{(l+1)\theta} 
F_l\left( (x_c-x) \epsilon^{-\phi}\right).
\end{equation}
We show that the order $\epsilon^{(l+2)\theta}$ term in the expansion
of the functional equation yields a linear, inhomogeneous differential 
equation of first order for the correction-to-scaling function $F_l(s)$.
First note that, due to (\ref{form:qshift2}), the expansion of the
$q$-algebraic functional equation contains terms of the form
\begin{equation}
\epsilon^{r_1} \left( s\epsilon^\phi \right)^{r_2}
\prod_{i,j} \left( \epsilon^{(l_i+1)\theta+m_j -n_j\phi} 
F_{l_i}^{(k_j)} s^{m_j-n_j}\right)^{r_{ij}} 
\end{equation}
with non-negative integers $r_1$, $r_2$, $r_{ij}$.
Here, $F_l^{(k)}$ denotes the $k$-th derivative of the correction-to-scaling function 
$F_l(s)$, and we have the constraints $0\le n_j \le k_j \le m_j$.
Let us analyze the order $\epsilon^{(l+2)\theta}$ in detail.
It consists of terms quite similar to the expansion in
(\ref{form:scaldglsqr}).
The contributions from the highest correction-to-scaling functions are
\begin{equation}\label{form:deq}
\left( \sum_k \partial_k P \right) F_{l+1}(s)+
x_c \left(\sum_k k \partial_k P \right) F_l'(s) + 
\left( \sum_{k,l} \partial_{k,l} P \right) F_l(s) F_0(s).
\end{equation}
Since the first prefactor vanishes by the assumption of a branch point, $F_{l+1}(s)$ does
not contribute to that order.
There are however non-vanishing contributions linear in $F_l(s)$ and its derivative.
The remaining terms contain products of the scaling function and corrections to scaling 
of lower order than $l$ together with their derivatives.
Given the functions $F_k(s)$ for $k<l$ and the desired asymptotic
behaviour as discussed in the previous paragraph, integration of the above
equation uniquely determines $F_l(s)$.

We note that for special solvable cases like staircase polygons, the corrections to 
scaling have a simple form.
They can be written as a polynomial in the scaling function and the scaling variable.
This is demonstrated below.

\subsection{Scaling function about a simple pole}

The simple pole is characterized by
\begin{equation}
a_M(x_c) = 0, \qquad a_{M}'(x_c) \neq 0, \qquad a_{M-1}(x_c) \neq 0. 
\end{equation}
About a simple pole, the perimeter generating function is diverging.
This means that there is no regular term in (\ref{form:scalbeh}) and that the exponent
$\theta$ is negative.
Let us therefore write the leading behaviour about the critical point as
\begin{equation}
G(x,q) \sim (1-q)^{-\theta} F_0( (x_c-x) (1-q)^{-\phi}) 
\qquad (x,q) \to (x_c^-,1^-).
\end{equation}
Note that we have redefined $\theta$ such that it is positive.
Since we are dealing with a simple pole, we can assume $\theta=\phi$.
We now derive a differential equation for the scaling function $F_0(s)$.
This is done by expanding the functional equation about $s=(x_c-x)\epsilon^{-\phi}$ and
taking the limit $\epsilon\to 0$.
To this end, we first multiply the functional equation by the exponent of its leading
divergence $\epsilon^{M\theta}$, where $M$ is the degree of the functional equation.
To lowest order in $\epsilon$, we get contributions from order $M$ and order $M-1$ of
the form
\begin{equation}
\begin{split}
0 = -a_M' s F_0(s)^M & \epsilon^{\phi} + 
\left( \sum_{{\bf k}_M}|{\bf k}_M| a_{{\bf k}_M}\right) x_c F_0'(s) F_0(s)^{M-1}
\epsilon^{1-\phi} + \\
& + a_{M-1} F_0(s)^{M-1}\epsilon^\theta + \ldots
\end{split}
\end{equation}
In order to obtain a non-trivial scaling equation, we demand equality of the
three expressions and get exponents
\begin{equation}
\theta=\frac{1}{2}, \qquad \phi=\frac{1}{2}.
\end{equation}
These exponents follow already from the analysis of the area weighted moments of the
previous section.
Here, their values arise from a consistency argument.
Assuming that the second term in the above expansion does not vanish, 
we get a differential equation of the form
\begin{equation}
s F_0(s) + \frac{f_{1,0}}{f_{0,0}} F_0'(s) - f_{0,0} = 0,
\end{equation}
where the coefficients $f_{1,0}$ and $f_{0,0}$ are given by 
(\ref{form:pf0}) and (\ref{form:pf1}).
The solution of this equation is uniquely determined by its prescribed
asymptotic behaviour and is given by
\begin{equation}
F_0(s) = f_{0,0} \sqrt{\pi a} \, \mbox{erfc}\left(\sqrt{a} s \right) e^{a s^2},
\end{equation} 
where he have introduced the abbreviation $a=-f_{0,0}/(2 f_{1,0})$, and 
$\mbox{erfc}(x) = 2/\sqrt{\pi} \int_x^\infty e^{-t^2}\, dt$ is the complementary error
function.
The coefficients $f_{n,0}$ in the asymptotic expansion of the scaling function are
exactly those derived in the previous section.
The asymptotic behaviour is given by
\begin{eqnarray}
F_0(s) &\sim& f_{0,0} s^{-1} \qquad (s\to \infty)\\
F_0(0) &=& f_{0,0}\sqrt{\pi a}\nonumber\\
F_0(s) &\sim& 2 F_0(0) \, e^{as^2} \qquad (s\to -\infty) \nonumber
\end{eqnarray}
The first singularity of the scaling function on the negative real
axis is at minus infinity, which implies $x_c(1)>x_c$ for the critical line
$x_c(q)$\mycite{PO95a}.

\subsection{Corrections to scaling about a simple pole}
The analysis of the singular behaviour of the area weighted moments
justifies the formal scaling Ansatz
\begin{equation}
G(x,q) = \sum_{l=0}^\infty \epsilon^{(l-1)\theta} 
F_l \left( (x_c-x) \epsilon^{-\phi} \right).
\end{equation}
As in the previous paragraph, we multiply the functional equation by
$\epsilon^{M\theta}$ in order to obtain a finite limit as $\epsilon\to 0$.
We show that the order $\epsilon^{(l+1)\theta}$ term in the expansion
of the functional equation yields a linear, inhomogeneous differential 
equation of first order for the correction to scaling function $F_l(s)$.
This is done in analogy to the case of a square-root singularity
discussed above.
Note first that the expansion of the $q$-algebraic functional equation leads to terms 
of the form
\begin{equation}
\epsilon^{M\theta} \epsilon^{r_1} \left( s\epsilon^\phi \right)^{r_2}
\prod_{i,j} \left( \epsilon^{(l_i-1)\theta+m_j -n_j\phi} 
F_{l_i}^{(k_j)} s^{m_j-n_j}\right)^{r_{ij}} 
\end{equation}
with non-negative integers $r_1$, $r_2$, $r_{ij}$.
We have the constraints $0 \le n_j \le k_j \le m_j$.
Let us analyze the order $\epsilon^{(l+1)\theta}$ term in detail.
The terms with highest corrections to scaling are
\begin{equation}
\begin{split}
\left( \sum_k P_k \right) F_{l+1}(s)& F_0(s)^{M-1}
+ x_c  \left( \sum_{{\bf k}_M}|{\bf k}_M| a_{{\bf k}_M}\right) F_{l}'(s) F_0(s)^{M-1}\\
& - 2 a_M' s F_{l}(s) F_0(s)^{M-1}
\end{split}
\end{equation}
The first term does not contribute since the prefactor vanishes by the assumption of a 
branch point.
The following terms contain $F_l(s)$ and its derivative linearly.
The remaining terms contain products of the scaling function and correction-to-scaling 
functions of lower order than $l$ together with their derivatives.
Given the functions $F_k(s)$ with $k<l$ and the desired asymptotic
behaviour as discussed in the previous paragraph, integration of the above
equation uniquely determines $F_l(s)$.

For simple polygon models such as Ferrers diagrams, the correction-to-scaling
functions can be written as a polynomial in the scaling function and the 
scaling variable, as we will see below.

\section{Numerical methods}

We now explain our methods for extracting scaling behaviour of polygon
models from given series data.
The first method involves a numerical analysis of the singular
behaviour of the area weighted moments by a fit of series enumeration data
to the expected asymptotic form.
This may then be compared to predictions about the scaling function
and corrections to scaling.
Another independent method consists in the computation of $q$-algebraic approximants.
Here, series enumeration data is used to compute an approximate $q$-algebraic
functional equation.
For polygon models, we believe this to be a good approximation about the critical
point, since the class of $q$-algebraic functions is rather general and 
since the scaling function seems to be determined by the nature of the
singularity of the perimeter generating function.
The method will then result in approximate differential equations for the 
scaling functions and will provide accurate estimates of its parameters.
We focus here on algebraic singularities, which is expressed in the requirement 
that the perimeter generating function shall be approximated by algebraic 
functions\mycite{BG90a}.

\subsection{Moment analysis}

We show how to estimate amplitudes of the area weighted moments
from given series enumeration data.
This in turn provides an estimate of the scaling function and correction-to-scaling 
functions, as follows from paragraph (2.4).

The coefficients $f_{k,l}$ of the area weighted moments
(\ref{form:momex}) determine the asymptotic growth of the Taylor coefficients of 
the functions $g_k(x)$.
This can be seen from the fact that asymptotically
\begin{equation}
\begin{split}
[x^n] & \frac{1}{(x_c-x)^\gamma} =\\
& \frac{n^{\gamma-1}}{x_c^n} 
\left( 1 + \frac{\gamma(\gamma-1)}{2n} +
\frac{\gamma(\gamma-1)(\gamma-2)(3\gamma-1)}{24 n^2} 
+ {\cal O}(n^{-3})\right).
\end{split}
\end{equation}
The analysis of series data is typically done by a direct fit to the expected 
asymptotic form\mycite{CG96, JG99}.
This provides estimates of amplitudes $A_{k,l}$ 
\begin{equation}
[x^n] g_k(x) \simeq x_c^{-n} n^{\gamma_{k,0}-1}\left( A_{k,0} + 
\frac{A_{k,1}}{n^{e(1)}} + \frac{A_{k,2}}{n^{e(2)}} + \ldots \right) 
\end{equation}
which can be used to obtain estimates of the coefficients $f_{k,l}$.
In particular, we have
\begin{equation}\label{form:numamp}
A_{k,0} = \frac{f_{k,0}}{x_c^{\gamma_{k,0}}\Gamma(\gamma_{k,0})}.
\end{equation}
This expresses the coefficients in the asymptotic expansion of the scaling function with
the leading asymptotic behaviour of the area moment expansion.
The subsequent amplitudes determine the functions describing the corrections to scaling.
For these amplitudes, the contributions from each term in (\ref{form:momex}) mix and have
to be analyzed carefully.
The usual procedure is to {\it assume} a special sequence of exponents
$e(1), e(2), \ldots$ and to analyze the rate of convergence of the approximation.
If exponents are omitted, then the coefficients in the approximation will tend to
infinity.
If assumed exponents are not present, their estimated amplitude will tend to zero.
A more detailed exposition of the method may be found in\mycite{CG96,JG99}.   

\subsection{$q$-algebraic approximants}

The idea of $q$-algebraic approximants is to fit a $q$-algebraic functional equation
to a given function $G(x,q)$
\begin{equation} \label{form:algapp}
P(y_0,\ldots,y_N,x,q) = \sum_{m=1}^M \sum_{{\bf k}_m} 
a_{{\bf k}_m}(x,q) y_{k_1} \cdots y_{k_m} + a_0(x,q) = 0
\end{equation}
such that (\ref{form:algapp}) is {\it exact} up to a given degree in the variables.
We restrict $a_{{\bf k}_m}(x,q)$ and $a_0(x,q)$ to be polynomials in $x$ and $q$ of
degrees not exceeding $d_x$ and $d_q$, respectively.
The coefficients of the polynomials can be found by solving the system of linear
equations deriving from the expansion of (\ref{form:algapp}) in its variables.
This process is not unique as there are different possibilities to order the set of
equations.
We demand that the limit $q \to 1$ of the approximation leads to {\it algebraic}
approximants\mycite{BG90a} to the perimeter generating function $G(x,1)$.
This is achieved as follows.
We fix a maximal degree $n_x$.
We expand (\ref{form:algapp}) about $x=0$ and $q=1$ up to the maximal degree $n_x$.
This yields a finite number of equations.
We order this set of equations in increasing powers of $1-q$ and, for each power, by
increasing powers of $x$.
The approximant is obtained by taking, for each power of $1-q$, as many equations as 
possible to obtain a nontrivial solution of the resulting system.
This gives the $q$-algebraic approximant of degree $d_x$ and $d_q$.
In the limit $q\to 1$, it reduces to the algebraic approximant of degree $d_x$ in $x$. 
If the class of $q$-algebraic approximant is correctly chosen, the approximate scaling
functions will converge to the correct scaling behaviour with increasing precision of
approximation $d_x$ and $d_q$.

$q$-linear approximants have been introduced recently\mycite{RG01}, 
together with a discussion of the relation between this new type of approximants 
and the existing method of partial differential approximants\mycite{Sty90}.
$q$-linear approxiamnts of first order are suited for the analysis of scaling behaviour 
about poles.
They led to the detection of the exact scaling functions for exactly solvable $q$-linear 
polygon models with a complicated functional equation of higher order\mycite{RG01}.
We will demonstrate our approach for $q$-quadratic approximants.

We have shown above that a square-root singularity leads to a scaling function of
Riccati type.
Since this behaviour is already exhibited by a $q$-algebraic functional equation of
degree two, $q$-quadratic approximants of first order may be employed to approximate 
such a scaling function.
Note however that the corrections to scaling may not be correctly approximated since
contributions from higher degrees of the functional equation enter here.
As a $q$-quadratic approximant we choose
\begin{equation}\label{form:qquad}
P(y_0,y_1,x,q) = y_0^2 + a(x,q) y_0 y_1 - b(x,q) y_0 + c(x,q).
\end{equation}
From the considerations above, the critical point $x_c$ is given by
\begin{equation}
b^2(x_c,1) = 4 c(x_c,1) \left( 1+ a(x_c,1) \right).
\end{equation}
The coefficients of the differential equation for
the scaling function are
\begin{eqnarray}
f_{0,0}^2 &=& \frac{a'b^2}{4(1+a)^3}-\frac{b'b}{2(1+a)^2} + \frac{c'}{(1+a)}, \\
-4f_{1,0} &=& x_c\frac{ab}{2(1+a)^2}, \nonumber
\end{eqnarray}
where the prime denotes differentiation, and all functions are evaluated at
$(x,q)=(x_c,1)$.
We will see below that this choice of approximant converges well for column-convex
polygons and rooted SAPs, which both have a square-root singularity in their perimeter
generating function.

\section{Examples}

We first discuss exactly solvable examples, which illustrate the general considerations
above.
These are rectangles, Ferrers diagrams and staircase polygons.
Two other exactly solvable examples are convex polygons and column-convex polygons,
which obey complicated $q$-algebraic functional equations.
We use these models in order to test our numerical tools.
We finally analyze the model of self-avoiding polygons numerically.

\subsection{Rectangles}

The perimeter and area generating function of rectangles satisfies the
$q$-linear functional equation
\begin{equation}
P(y_0,y_1,x,q) = qx^2 (1-qx) y_1 - (1-qx) y_0 + q x^2(1+qx) = 0.
\end{equation}
It can be derived from the expression given previously\mycite{PO95a} for the anisotropic perimeter
and area generating function by symmetrization or by a graphical
argument similar to that given previously\mycite{PO95a}.
The perimeter generating function is rational with a double pole at the critical point
$(x,q)=(1,1)$.
Using an argument analogous to the one given above for a simple pole, we assume that
the perimeter and area generating function admits an expansion about the critical
point of the form
\begin{equation}
G(x,q) = \epsilon^{-1} \sum_{k=0}^\infty \epsilon^{\frac{k}{2}} F_k
\left(s\epsilon^{-\frac{1}{2}} \right),
\end{equation}
where $s=x_c-x$ and $\epsilon=1-q$.
It is seen by arguments analogous to those above that the expansion of the 
$q$-linear functional equation yields differential equations for the functions $F_k(s)$. 
The first few functions are given by
\begin{eqnarray}
F_0(s) &=& \mbox{Ei}(s^2) e^{s^2}\\
F_1(s) &=& s^3 F_0(s) -s-s^{-1} \nonumber\\
F_2(s) &=& \left( \frac{1}{2} s^6 + \frac{11}{12} s^4 -\frac{1}{2} s^2
-\frac{1}{2} \right)F_0(s) -\frac{1}{2}s^4-\frac{5}{12}s^2+\frac{17}{12} +\frac{1}{6}
s^{-2} \nonumber\\
F_3(s) &=& \left(\frac{1}{6} s^9 +\frac{11}{12}s^7+\frac{1}{3}s^5-s^3 \right) F_0(s)
-\frac{1}{6} s^7-\frac{3}{4}s^5+ \nonumber \\
&& +\frac{1}{4} s^3+\frac{1}{2} s-\frac{1}{6}s^{-1}, \nonumber
\end{eqnarray}
where $\mbox{Ei}(x) = \int_{1}^\infty e^{-tx}/t \, dt$ denotes the exponential integral.
The coefficients $f_{n,0}$ in the asymptotic expansion of the scaling function are given
by $f_{n,0}=(-1)^n n!$.
All corrections to scaling appear to be linear in $F_0(s)$.
Despite the simple form of the functional equation, a recurrence for the coefficients
is not transparent.
We note in passing that the scaling function is singular at the
origin.
This is possibly related to the fact that the condition (\ref{form:p2}) 
is violated for rectangles.

\subsection{Ferrers diagrams}

The perimeter and area generating function of Ferrers diagrams satisfies the
$q$-linear functional equation
\begin{equation}
P(y_0,y_1,x,q) = q x^2 y_1 - (1-qx)^2 y_0 + q x^2 =0.
\end{equation}
This equation can be derived from the functional equation for the aniso\-tropic model
given previously\mycite{PO95a} by symmetrization or, alternatively, by a graphical argument
similar to that given there.
The perimeter generating function is rational with a simple pole at $x=1/2$.
Thus the considerations of the previous section apply.
About the critical point $(x,q)=(1/2,1)$ we assume an expansion of the
perimeter and area generating function of the form
\begin{equation}
G(x,q) = \epsilon^{-\frac{1}{2}} \sum_{k=0}^\infty \epsilon^{\frac{k}{2}} F_k
\left(s\epsilon^{-\frac{1}{2}} \right).
\end{equation}
The scaling function $F_0(s)$ has first been given previously\mycite{RG01} and 
has also been extracted for a semi-continuous version of the model\mycite{PO95a}.
The first few functions $F_k(s)$, as computed from the functional equation, are
\begin{eqnarray}
F_0(s) &=& \sqrt{\frac{\pi}{8}} \, \mbox{erfc}( \sqrt{8} s)  e^{8s^2}\\
F_1(s) &=& \left( \frac{32}{3} s^3 + 2s \right) F_0(s) -\frac{2}{3}-\frac{4}{3}s^2 
\nonumber\\
F_2(s) &=& \left( \frac{512}{9}s^6 +\frac{128}{3}s^4-4s^2-\frac{1}{24} \right) F_0(s)
-\frac{64}{9}s^5-\frac{44}{9} s^3 +\frac{5}{4}s \nonumber\\
F_3(s) &=& \left( \frac{16384}{81}s^9+\frac{1024}{3}s^7 +
\frac{512}{15}s^5 -\frac{124}{9}s^3 -\frac{3}{4}s \right) F_0(s) \nonumber \\
 && -\frac{2048}{81} s^8 -\frac{3328}{81}s^6-\frac{256}{135}s^4+\frac{427}{270}s^2+
\frac{7}{135},\nonumber
\end{eqnarray}
where $\mbox{erfc}(x)$ is the complementary error-function.
All corrections to scaling are linear in the scaling function, with polynomial
coefficients in the scaling variable $s$.
A simple recursion for these polynomials is not apparent.

\subsection{Staircase polygons}

The perimeter and area generating function of staircase polygons satisfies the
$q$-quadratic functional equation
\begin{equation}
P(y_0,y_1,x,q) = y_1 y_0 + (2xq-1)y_0 + x^2q = 0.
\end{equation}
It can be derived from the functional equation for the anisotropic perimeter and area
generating function given previously\mycite{PB95} by symmetrization.
Alternatively, it can be derived from a concatenation argument similar
to that given there.
The perimeter generating function has a square-root singularity at $x_c=1/4$, such that
the considerations of the previous section apply.
About the critical point $(x,q)=(1/4,1)$, we assume an expansion of
the perimeter and area generating function of the form
\begin{equation}
G(x,q) = G^{(reg)}(x,q) + \sum_{k=0}^\infty \epsilon^{\frac{k+1}{3}} F_k
\left(s\epsilon^{-\frac{2}{3}} \right),
\end{equation}
where $s=x_c-x$ and $\epsilon=1-q$.
The scaling function $F_0(s)$ and the correction-to-scaling functions $F_k(s)$, where
$k>0$, can be computed recursively from the expansion of the
functional equation.
The scaling function $F_0(s)$ was first computed\mycite{P95}
using an explicit expression for the perimeter and area generating
function.
It was recently extracted from the $q$-functional equation\mycite{RGJ01}.
The first few contributions $F_k(s)$ are
\begin{eqnarray}
F_0(s) &=& \frac{1}{16} \frac{d}{ds} \log \mbox{Ai} \left( 2^{\frac{8}{3}} s\right)\\
F_1(s) &=& 0 \nonumber\\
F_2(s) &=& \frac{3}{40} -\frac{16}{15}s F_0(s)+ \frac{32}{15}s^2 F_0'(s)\nonumber\\
F_3(s) &=& \frac{1}{6} F_0(s) -\frac{1}{3}s F_0'(s)\nonumber\\
F_4(s) &=& -\frac{89}{350}s-\frac{1936}{1575}s^2 F_0(s) + \left(\frac{1888}{525}s^3 +
\frac{9}{2240} \right) F_0'(s) + \nonumber\\
&&+ \frac{512}{225}s^4 F_0''(s)\nonumber
\end{eqnarray}
Note that the first correction to scaling vanishes.
The above expressions suggest that all corrections to scaling may be expressed in terms 
of derivatives of the scaling function with polynomial coefficients.
A simple recursion for the polynomials is not transparent.

\subsection{Convex polygons}

The anisotropic perimeter and area generating function $G(x,y,q) = C(x)$ 
of convex polygons\mycite{L91, Bou92} satisfies the following system of 
$q$-algebraic equations\mycite{Bou92}
\begin{equation}
\begin{split}
A(x) &= \frac{xyq}{1-xq} + \frac{y+A}{1-xq}A(xq) \\
\begin{pmatrix} B\\ B_1\end{pmatrix} (x) & =  \frac{xyq}{1-xq} 
\begin{pmatrix}1 \\ 1 \end{pmatrix} + 
\frac{y+A}{1-xq}\begin{pmatrix}1 & xq \\ 1 & 1\end{pmatrix}
\begin{pmatrix} B \\ B_1\end{pmatrix} (xq)
\\
\begin{pmatrix} C\\ C_1 \\ C_2 \end{pmatrix} (x) 
& =  
\frac{xyq}{1-xq} \begin{pmatrix} 1 \\ 1 \\1 \end{pmatrix} + 
\\
+\frac{2}{y+A} & \begin{pmatrix} V^2 \\ VW \\ W^2 \end{pmatrix}
\frac{y}{(1-xq)^2}\begin{pmatrix} 1 & 2 xq & x^2 q^2  \\ 1 & 1+ xq & xq \\ 1
& 2& 1\end{pmatrix} \begin{pmatrix} C \\ C_1 \\ C_2\end{pmatrix} (xq),
\end{split}
\end{equation}
where the functions $V$ and $W$ are given by
\begin{equation}
V = B - \frac{xyq}{1-xq}, \qquad W = B_1 - \frac{xyq}{1-xq}.
\end{equation}
The function $A(x)$ is the anisotropic perimeter and area generating function for
staircase polygons.
This system of $q$-algebraic equations can be reduced to a single $q$-algebraic
functional equation involving only $G(x,y,q)$:
Shifting the parameter $x$ to $qx$ in the above system gives five new equations, but
only four new variables $A(q^2 x)$, $B(q^2 x)$, $B_1(q^2 x)$, $C_1(q^2 x)$ and $C_2(q^2
x)$.
This process is repeated as long as the number of variables remains
larger than the number of equations.
The variables in this system of equations can then all be eliminated,
leading to a $q$-algebraic functional equation for the anisotropic
generating function $G(x,y,q)$.
Since the generating function is symmetric, a $q$-algebraic generating function for
the isotropic model $G(t,t,q)$ can in principle be derived from that of the 
anisotropic model by symmetrization.
It is however very complicated.

We restrict ourselves to the perimeter and area generating function of the isotropic 
model in the following.
We used the above system to generate series data for the anisotropic perimeter and area
generating function.
We analyzed the isotropic case $y=x$ and observed that the area moments
$g_0(x)$ to $g_7(x)$ satisfy an algebraic equation of degree four. 
The perimeter generating function\mycite{Bou92} is given by
\begin{equation}
G(x,1) = \frac{x^2(1-6x+11x^2-4x^3)}{(1-4x)^2} - \frac{4x^4}{(1-4x)^{3/2}}.
\end{equation}
It has a singularity at $x=1/4$ with a double pole as leading singularity.
We conjecture that the singular behaviour of all area moments is of the form
\begin{eqnarray}
g_{2n}(x) &=& \frac{r_{2n}(x)}{(1-4x)^{4n+2}} + \frac{s_{2n}(x)}{(1-4x)^{3n+3/2}} \\
g_{2n+1}(x) &=& \frac{r_{2n+1}(x)}{(1-4x)^{4n+4}} +
\frac{s_{2n+1}(x)}{(1-4x)^{3n+5/2}}, \nonumber 
\end{eqnarray}
where the polynomials $r_n(x)$ and $s_n(x)$ are non-zero at $x=x_c$.
The leading singular terms give rise to an asymptotic expansion of the scaling function
\begin{equation}
\begin{split}
F_0(s)
= & \frac{1}{2048} s^{-2} -\frac{1}{65536} s^{-4} +\frac{1}{1048576} s^{-6}
-\frac{3}{33554432} s^{-8} \\
& +\frac{3}{268435456} s^{-10} 
 -\frac{15}{8589934592} s^{-12} +\frac{45}{137438953472} s^{-14}\\
& -\frac{315}{4398046511104} s^{-16} + {\cal O}\left( s^{-18} \right),
\end{split}
\end{equation}
where $s=x_c-x$.
This we recognize as
\begin{equation}
F_0(s) = \frac{e^{32 s^2}}{64} \mbox{Ei}(32 s^2),
\end{equation} 
where $\mbox{Ei}(s)$ is the exponential integral.
The scaling function for convex polygons is the same as for rectangles.
This means that convex polygons scale in a first approximation like rectangles.
The first correction to scaling due to the simple pole is linear in the scaling
function, as in the rectangle case.
We have not found a closed expression for the leading corrections to scaling due to the
square-root singularity.

\subsection{Column-convex polygons}

The perimeter and area generating function of column-convex polygons $G(x,y,q)$
satisfies the following system of $q$-quadratic functional equations\mycite{BOP94}
\begin{equation}\label{form:colcon}
\begin{split}
A(x) =& (1+B(qx)-A(qx))(qxy+qxB(x)) + y A(qx) + 2 A(x) A(qx) \\
B(x) =& A(x) + (C(qx)-B(qx)) qx (y+B(x)) + (y+2A(x)) B(qx) \\
C(x) =& B(x) + (C(qx)-B(qx)) (qxy+qx C(x)) + \\
& +(y+2B(x))B(qx)+yC(qx),
\end{split}
\end{equation}
where $G(x,y,q)=A(x)$. 
This system of $q$-algebraic equations can be reduced to a single $q$-algebraic
functional equation involving only $G(x,y,q)$:
Shifting the parameter $x$ to $qx$ in the above system gives three new equations, but
only two new variables $B(q^2 x)$, $C(q^2 x)$.
This process is repeated as long as the number of variables remains
larger than the number of equations.
The variables can then all be eliminated, leading to a $q$-algebraic functional equation
for the anisotropic generating function $G(x,y,q)$.

We consider the behaviour of the isotropic perimeter and area generating function
$G(t,q)=G(t,t,q)$.
Since the generating function is not symmetric in $x$ and $y$, we
cannot conclude from the above argument that the isotropic generating
function $G(t,q)$ is $q$-algebraic as well.
The limit $q \to 1$ in (\ref{form:colcon}) yields an algebraic
equation of degree four for the perimeter generating function\mycite{BOP94}.
Its singularity of smallest modulus is of square-root type at $t_c=3-2\sqrt{2}$.
Assuming an underlying $q$-algebraic functional equation, 
we can derive the exact scaling function about $(t,q)=(t_c,1)$ 
by extracting the leading singular behaviour of the perimeter
generating function and its first area weighted moment%
\footnote{The scaling function for a semicontinuous version of the 
model has been derived earlier\mycite{BOP94}.}.
These two functions satisfy algebraic equations of degree four\mycite{BOP94}.
The amplitude $f_0^2=0.13288910\ldots$ is given as a root of the quartic equation
\begin{equation}
4879681 t^4 + 52908512 t^3 - 6131584 t^2 - 161792 t + 4096 = 0.
\end{equation}
The amplitude $f_1=-0.0035579552\ldots$ is given as a root of the quartic equation
\begin{equation}
565504 t^4 -1580288 t^3 +216672 t^2 +1072 t + 1 = 0.
\end{equation}
Assuming that the scaling function is given by (\ref{form:logairy}), we can compare the
predictions for the leading singular behaviour of the area moments with numerical data
from series extrapolation.
More precisely, we compare the amplitudes obtained by numerical
analysis with those predicted by the scaling function (\ref{form:logairy}).
We therefore generated the series expansion of the first ten area weighted moments from
the functional equation up to perimeter 40 and used the method described above.
We assumed corrections to the asymptotic behaviour with half-integer
exponents, $e(m)=m/2$, and performed an approximation with ten correction terms.
The amplitude estimates are shown in the table below. 

\vspace{2ex}

\centerline{
\begin{tabular}{|c||r|r|r|}\hline
$k$ & 
\multicolumn{1}{c|}{$f_{k,0}$} & 
\multicolumn{1}{c|}{$f_{k,1}$} & 
\multicolumn{1}{c|}{$f_{k,2}$}\\
\hline \hline
0 & 
-0.3645(1) $\times 10^{-2}$&
0.0000(1) $\times 10^{-2}$& 
0.234(1) $\times 10^{-1}$\\
1 & 
-0.3557(1) $\times 10^{-2}$&  
0.7557(1) $\times 10^{-2}$& 
0.000(1) $\times 10^{-2}$\\
2 &  
0.8681(1) $\times 10^{-4}$& 
-0.1475(1) $\times 10^{-3}$& 
-0.855(1) $\times 10^{-3}$\\
3 & 
-0.5083(1) $\times 10^{-5}$&  
0.8999(1) $\times 10^{-5}$& 
0.757(1) $\times 10^{-4}$\\
4 &  
0.4569(1) $\times 10^{-6}$& 
-0.8432(1) $\times 10^{-6}$& 
-0.915(1) $\times 10^{-5}$\\
5 & 
-0.5472(1) $\times 10^{-7}$&  
0.1042(1) $\times 10^{-6}$& 
0.137(1) $\times 10^{-5}$\\
6 &  
0.8156(1) $\times 10^{-8}$& 
-0.1590(1) $\times 10^{-7}$& 
-0.247(1) $\times 10^{-6}$\\
7 & 
-0.1452(1) $\times 10^{-8}$&  
0.2882(1) $\times 10^{-8}$& 
0.516(1) $\times 10^{-7}$\\
8 &  
0.3006(1) $\times 10^{-9}$& 
-0.6037(1) $\times 10^{-9}$& 
-0.122(1) $\times 10^{-7}$\\
9 & 
-0.7093(1) $\times 10^{-10}$& 
0.1422(1) $\times 10^{-9}$& 
0.330(1) $\times 10^{-8}$\\
10 & 
0.1877(1) $\times 10^{-10}$&
-0.3677(1) $\times 10^{-10}$& 
-0.997(1) $\times 10^{-9}$\\
\hline
\end{tabular}
}

\vspace{2ex}

The values $f_{0,1}$ and $f_{1,2}$ are zero up to numerical accuracy.
They corresond to non-negative integer exponents, which cannot be
extracted from finite size extrapolation, since they are regular terms in
the corresponding generating function.
A comparison with the expressions $f_{k,0}$ computed from the assumed exact 
scaling function coincides with the data up to numerical accuracy.
We therefore assume that the expression given for the scaling function is correct.

In contrast to the model of staircase polygons, the first correction to scaling is
non-vanishing.
An analysis along the lines of that given in section (3.2) yields the
linear differential equation
\begin{equation}
F_1'(s) + c_0 F_1(s) F_0(s) = c_1 + c_2 s F_0(s) + c_3 F_0''(s)
\end{equation}
for the first correction to scaling $F_1(s)$ with constants $c_k$,
where $k=0,\ldots,3$.
We have in particular $c_0=-(2 f_{1,0})^{-1}$.
A numerical analysis now reveals that the constant $c_3$ is several
orders of magnitude smaller than $c_1, c_2$.
Assuming $c_3=0$, we can write solutions of the above equation in the form 
\begin{equation}
F_2(s) = a_0 s + a_1 F_0'(s)
\end{equation}
with constants $a_0, a_1$.
We may impose $a_0=0$, since any contribution may be absorbed in the
regular part.
The constant $a_1=-0.0414654242\ldots$ may be extracted from the
defining equations of the perimeter generating function and the area
moment.
It is given as a root of the equation
\begin{equation}
12032 t^4 - 67328 t^3 + 88288 t^2 + 8432 t + 193=0.
\end{equation}
Using this Ansatz, the values of the consecutive area moments are
predicted correctly within numerical accuracy.
We therefore assume that the above expression is correct.
It is in principle possible to derive algebraic equations for greater
moments of the area to check consistency.
This requires much higher computational effort, however.
Higher corrections to scaling may also be analyzed using the above
methods.
A numerical test is however more difficult to perform accurately
because the number of undetermined constants increases.

We also applied the method of $q$-quadratic approximants in order to obtain estimates for the
amplitudes $f_{0,0}$ and $f_{1,0}$.
To this end, we computed approximants of the form (\ref{form:qquad}) to the series data
for polynomials of degree one in $\epsilon$ and up to degree 9 in $t$.
We get $f_{0,0}^2=0.1328(1)$ and $f_{1,0}=-0.003557(1)$, agreeing with the exact
values up to numerical accuracy.

\subsection{Self-avoiding polygons}

The problem of self-avoiding polygons\mycite{LSF87, F89, FGW91} is an
as yet unsolved combinatorial problem.
The asymptotic form of the finite-size scaling function in the regime
$q>1$ has been determined recently\mycite{PO99}.
It has recently been proved that the anisotropic perimeter generating function is not
$D$-finite\mycite{R00}.
This suggests that the isotropic perimeter generating function has the same property.
If so, it cannot then be algebraic.

Extensive numerical analysis of series data has shown that the
perimeter generating function has a finite singularity with exponent
$3/2$ at the critical point $x_c=0.143680628(1)$\mycite{FGW91, J00}.
This means that the perimeter and area generating function for {\it rooted} SAPs 
$G^{(r)}(x,q)=x\frac{d}{dx}G(x,q)$ has a square-root singularity.
About the critical point, numerical analysis of series data\mycite{J00} strongly
supports an expansion of the form (\ref{form:sqroot}) with exponents
increasing by one (instead of $1/2$).
We previously analyzed series enumeration data for SAPs on the
square-lattice, counted by perimeter and area, and extracted the
functional form of the scaling function\mycite{RGJ01}.
Here, we explain this analysis in more detail and extend it to the 
first two corrections to scaling.

We analyzed series enumeration data of the first ten area-weighted
moments of the half-perimeter generating function for rooted SAPs.
The data was produced up to half-perimeter 43 using the finite-lattice
method and is available on request\mycite{RGJ01}.
We found the same exponents $\phi=2/3$ and $\theta=1/3$ as for
staircase polygons, confirming a previous investigation\mycite{PO95b}.
We then assumed corrections to the asymptotic behaviour with
half-integer exponents, $e(m)=m/2$, and performed an approximation 
with ten correction terms.
The first few values of the amplitudes $f_{k,0}, f_{k,2}, f_{k,3}$, 
obtained by an analysis similar to that described in the previous
section, are shown in the table below.

\vspace{2ex}

\centerline{
\begin{tabular}{|c||r|r|r|}
\hline
$k$ & 
\multicolumn{1}{c|}{$f_{k,0}$} & 
\multicolumn{1}{c|}{$f_{k,2}$} & 
\multicolumn{1}{c|}{$f_{k,3}$}\\
\hline \hline
0 & 
-0.9296(1) $\times 10^{-0}$& 
-2.81(1) $\times 10^{-0}$& 
0.00(1) $\times 10^{-0}$
\\
1& 
-0.5716(1) $\times 10^{-2}$& 
 0.605(1) $\times 10^{-3}$&
-0.68(1) $\times 10^{-0}$\\
2 &
 0.8789(1) $\times 10^{-4}$& 
-0.799(1) $\times 10^{-3}$&
-0.65(1) $\times 10^{-2}$\\
3 & 
-0.3243(1) $\times 10^{-5}$&
 0.487(1) $\times 10^{-4}$&
 0.13(1) $\times 10^{-3}$\\
4 &
 0.1836(1) $\times 10^{-6}$&
-0.386(1) $\times 10^{-5}$&
-0.61(1) $\times 10^{-5}$\\
5 &
-0.1386(1) $\times 10^{-7}$&
 0.374(1) $\times 10^{-6}$&
 0.41(1) $\times 10^{-6}$\\
6 &
 0.1301(1) $\times 10^{-8}$&
-0.429(1) $\times 10^{-7}$&
-0.35(1) $\times 10^{-7}$\\
7 &
-0.1460(1) $\times 10^{-9}$&
 0.568(1) $\times 10^{-8}$&
 0.37(1) $\times 10^{-8}$\\
8 &
 0.1904(1) $\times 10^{-10}$&
-0.856(1) $\times 10^{-9}$&
-0.47(1) $\times 10^{-9}$\\
9 &
-0.2832(1) $\times 10^{-11}$&
 0.144(1) $\times 10^{-9}$&
 0.67(1) $\times 10^{-10}$\\
10 &
 0.4729(1) $\times 10^{-12}$&
-0.269(1) $\times 10^{-10}$& 
-0.10(1) $\times 10^{-10}$\\
\hline 
\end{tabular}
}

\vspace{2ex}

We also determined the coefficients $f_{k,1}$ for the first ten area
weighted moments.
They are all several orders of magnitude less than the amplitudes 
$f_{k,0}, f_{k,2}, f_{k,3}$.
Therefore, we assume that they are all zero and do not list them in
the table.
It has been argued by universality\mycite{C94} that
$f_{1,0}$ is given by $-x_c/(8\pi)$.
The universal amplitude combinations given previously\mycite{RGJ01} may be 
reconstructed from the values $f_{k,0}$ using (\ref{form:numamp}).

Though the perimeter generating function is almost certainly not
$D$-finite, it is still possible that rooted SAPs are described by a $q$-algebraic functional
equation, leading to an algebraic differential equation for the
perimeter generating function at some order $\epsilon^n$ with $n>0$.
Under this assumption, it is still possible to perform a scaling analysis along the
lines presented above.
In fact the argument is exactly the same as that presented above, so we conclude that we
might expect the differential equation (\ref{form:diffeq}) for the scaling function.
To test this assumption, we take the numerical estimates $f_{0,0}$ and
$f_{1,0}$ and compare the predictions for the amplitudes $f_{k,0}$ for
$k>1$ to the numerical estimates in the table according to (\ref{form:sqas1}).
The prediction coincides with the amplitudes for $k>1$ with an error less than $10^{-4}$.
We conclude that the assumed form of the scaling function
(\ref{form:logairy}) for rooted SAPs is correct.
Universality implies that the functional form of the scaling function
does not depend on the underlying lattice.
Therefore, the coefficients $c_n$ of (\ref{form:sqas1}) should be
obtainable  on any lattice from the amplitude combinations
\begin{equation}
c_n = f_{n,0} f_{0,0}^{n-1} f_{1,0}^{-n}= (-1)^n f_{n,0} f_{0,0}^{n-1}
\left(\frac{8\pi}{x_c} \right)^{n},
\end{equation}
up to normalization due to the given geometry.
This has been successfully demonstrated for the triangular lattice\mycite{RGJ01}.

It is also possible to apply the method of $q$-quadratic approximants
in order to obtain estimates of the constants $f_{0,0}$ and $f_{1,0}$.
We get $f_{0,0}=-0.9296(1)$ and $f_{1,0}= -0.005716(1)$.
These values coincide with the values obtained by moment analysis within numerical
accuracy given by the available series data.

We now analyze the corrections to scaling.
In order to obtain information about the first correction to scaling $F_1(s)$, we have
estimated the coefficients $f_{k,1}$ for the first ten area weighted moments.
As mentioned above, they are all several orders of magnitude less than
the terms $f_{k,0}$ and $f_{k,2}$.
We therefore conclude that the first correction is absent, that is $F_1(s)=0$.

Under the assumption that rooted SAPs satisfy a $q$-algebraic functional equation, power
counting and use of the differential equation for the scaling function results in a 
differential equation for the second correction to scaling $F_2(s)$.
It is given by
\begin{equation}
F_2'(s) + c_0 F_2(s) F_0(s) = c_1 s^2 + c_2 F_0(s) + c_3 s F_0^2(s) + c_4 F_0^4(s)
\end{equation}
with constants $c_k$, where $k=0,\ldots,4$.
Remarkably, the solution of this equation can be written as a linear combination of
derivatives of the scaling function 
\begin{equation}
F_2(s) = a_0 + a_1 s^3 + a_2 s F_0(s) + a_3 s^2 F_0'(s) + a_4 F_0''(s)
\end{equation}
with constants $a_k$, where $k=0,\ldots,4$.
We used this Ansatz for $F_2(s)$ and determined the constants $a_k$
from series enumeration data.
Note that knowledge of the constants $a_2$, $a_3$ and $a_4$ predicts the amplitudes of
all higher area moments.
Again, a numerical analysis reveals that the coefficient $a_4$ is several orders of
magnitude smaller than $a_2$ and $a_3$.
We therefore assume that it is equal to zero.
We extracted the numbers $a_2$ and $a_3$ by solving the Ansatz with series data at
order $s^{-3/2}$ and $s^{-3}$, yielding $a_2=1.01(1)$ and $a_3=4.01(1)$.
The other numbers are then given by $a_0=-0.01(1)$ and $a_1=0.00(1)$.
The first corrections to the area moments are then predicted with a relative error less
than $10^{-2}$.
Therefore, the above analysis leads to the conjecture
\begin{equation}
F_2(s) = s F_0(s) + 4 s^2 F_0'(s).
\end{equation}
It is difficult to analyze higher corrections to scaling since the
numerical accuracy in the determination of the amplitudes $f_{k,l}$
decreases with increasing $l$.

The scaling function and correction-to-scaling functions for SAPs can
be obtained from the corresponding functions for rooted SAPs by integration.

\section{Conclusion}

We have analyzed the singular behaviour of $q$-algebraic functional equations which
describe exactly solvable polygon models.
We considered the special case where the limit $q\to 1$ leads to an
algebraic equation.
We analyzed the two simplest types of singularity, a square-root singularity and
a simple pole.
It was observed that the scaling function obtained is universal -- it
does not depend on the degree of the functional equation.
We gave simple exactly solvable examples for each type of singularity.
It is obvious that the methods introduced may be used to analyze other types
of algebraic singularity in detail.
For example, a finite singularity of type $(x_c-x)^{1/n}$ leads to 
exponents $\phi=n/(2n-1)$ and $\theta=1/(2n-1)$, resulting in a 
differential equation for the scaling function of the 
form $F(s)^n + a F'(s) +b s =0$.
This generalizes the result for a square-root singularity and may be compared 
to exact scaling functions predicted recently\mycite{C01}.
A related question is which polygon models realize these more complicated
types of singularity?
A natural class to look at are models of interacting polymers, whose exactly solvable
members are again described by $q$-algebraic functional equations\mycite{OPB93}.

Although we discussed only the simplest forms of scaling behaviour,
this led to an apparently correct prediction for the unsolved model of 
self-avoiding polygons, where we
numerically confirmed predictions of the functional form of the
scaling function and its first two corrections to scaling.
We know however that anisotropic (rooted) self-avoiding polygons are not
$D$-finite\mycite{R00}.
This suggests that rooted self-avoiding polygons may have a
$q$-algebraic perimeter and area generating function, a conjecture
which would be interesting to consider more deeply.

A further step in the analysis of $q$-algebraic functional equations consists
in going beyond the algebraic case, resulting in a differentiably algebraic 
perimeter generating function.
Although the analysis in the most general case is difficult, since the singularity
structure of these types of functions is not fully understood, it is possible to analyze
at least the case of a linear differential equation in some detail.
This case arises from the $q\to 1$ limit of $q$-linear functional equations.
This may shed light on the scaling behaviour of polygon models such as three-choice
polygons\mycite{CGD97}, whose perimeter generating function diverges logarithmically.

From a mathematical point of view, the scaling behaviour of $q$-algebraic functional
equations provides examples of uniform asymptotic expansions.
There are only few cases known where the existence of such an
expansion has been proven, for example for staircase polygons\mycite{P95}, using
an explicit expression for the perimeter and area generating function. 
In the present paper, we extracted the scaling function and
corrections to scaling assuming the existence of such an expansion
and verified the results numerically.
It may be possible to prove existence directly from the defining
$q$-functional equation, which would answer the problem in a much more general framework.     

Self-avoiding polygons are related to a number of other systems via
the $O(n)$-mapping\mycite{V98}.
This poses the question whether the methods and results of scaling analysis
developed so far have applications to percolation and magnetic
systems, where only a few scaling results are available.

\section*{Acknowledgements}

The author thanks Iwan Jensen for providing series data for SAPs and Tony Guttmann for
help concerning numerical analysis, many fruitful discussions, 
and a careful reading of the manuscript.
The author has also benefitted from discussions with John Cardy, Mireille
Bousqet-M\'elou and Thomas Prellberg.
This work is supported by the German Research Council (DFG) and the Australian Research
Council (ARC).

\end{document}